Mark D. Hill and Vijay Janapa Reddi

# Viewpoint
# Accelerator-Level Parallelism

*Charging computer scientists to develop the science needed to best achieve the performance and cost goals of accelerator-level parallelism hardware and software.*

While past information technology (IT) advances have transformed society, future advances hold great additional promise. For example, we have only just begun to reap the changes from artificial intelligence—especially machine learning—with profound advances expected in medicine, science, education, commerce, and government. All too often forgotten, underlying the IT impact are the dramatic improvements in the programmable hardware. Hardware improvements deliver performance that unlocks new capabilities. However, unlike in the 1990s and early 2000s, tomorrow's performance aspirations must be achieved with much less technological advancement (Moore's Law and Dennard Scaling). How then does one deliver AR/VR, self-driving vehicles, and health wearables at costs that enable great customer value?

One approach that has emerged is to use **accelerators:** *hardware components that execute a targeted computation class faster and usually with much less energy*. An accelerator's flexibility can vary from high (GP-GPU) to low (fixed-function block). Recent work tends to focus on targeting specific application domains, such as graphics (before GPUs generalized), deep machine learning, physics simulations, and genomics. Moreover, most work on accelerators, including in articles appearing in *Communications*,[2,5,6] has focused on CPUs using a single accelerator, with one early forecast of multiple accelerator use.[1]

In our view, many future computing systems will obtain greater efficiency by employing multiple accelerators where each accelerator efficiently targets an aspect of the ongoing computation, much as a Swiss Army knife has specific tools for specific tasks. Smartphones foreshadow this future by employing many accelerators concurrently, but unlike a Swiss Army knife these accelerators often operate in parallel using separately developed software stacks.

We assert there is as yet no "science" for debating and systematically answering basic questions for how to best facilitate broad, flexible, and effec-

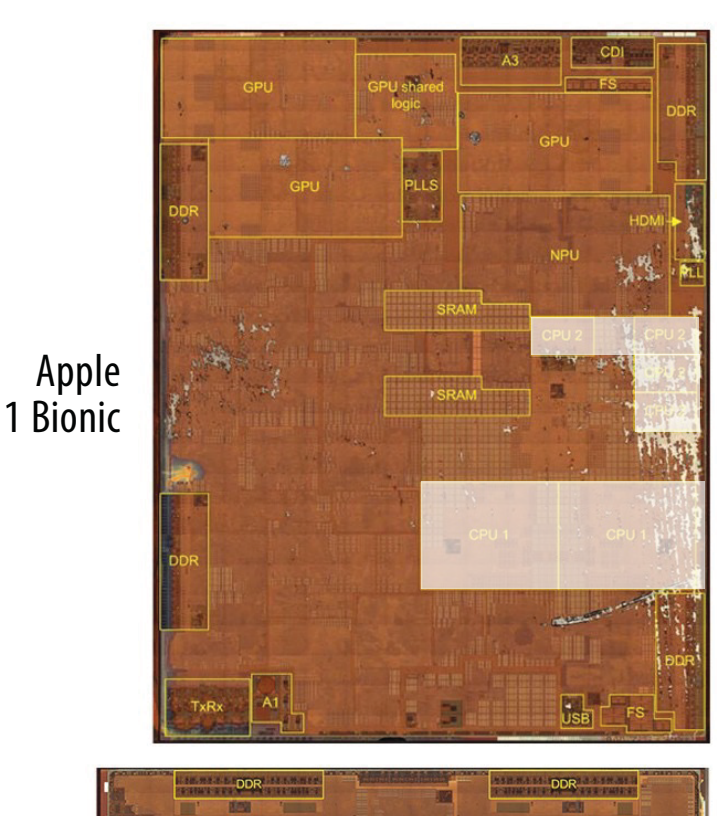

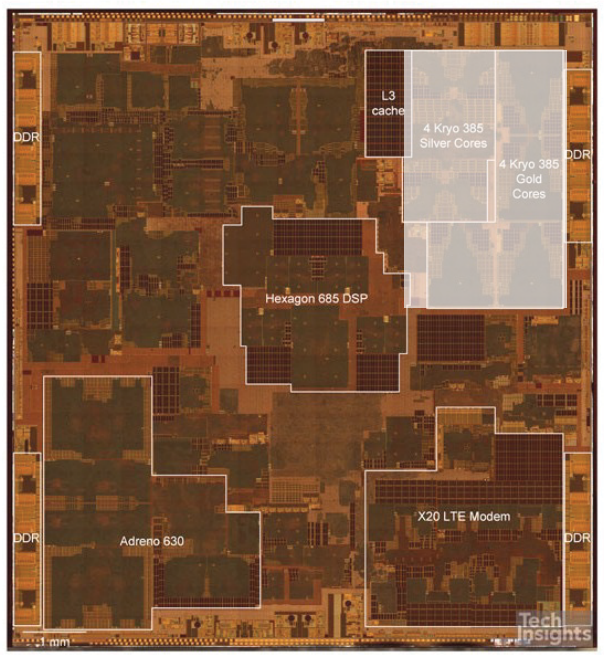

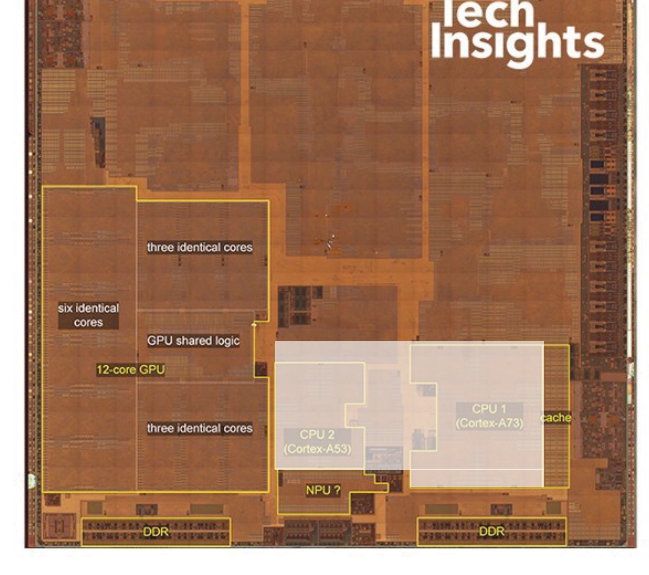

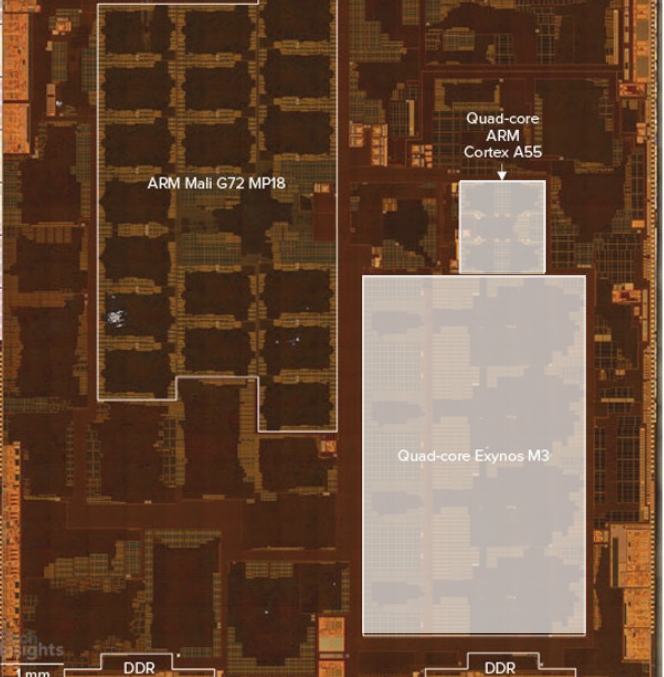

**Modern System-on-Chip (SoC) architectures. The CPUs in modern SoCs (shown in white) occupy only a small percentage of the die area. The rest of the SoC is committed to a potpourri of different accelerators, such as the DSP, GPU, ISP, NPU, video, and audio codecs.**

tive use of multiple accelerators. In this Viewpoint, we expose this opportunity (the what), but charge our readers with determining how best to address it. We review past computer system improvements exploiting levels of parallelism, and introduce Accelerator-Level Parallelism (ALP) as a way to frame new challenges, and expand on the "point" success of smartphone ALP.

## Past, Present, and Future Parallelism

As technology scaling provided more and smaller transistors, computer processor architects transformed the transistor bounty into faster processing by using the transistors in parallel. Effectively using repeated transistor doubling required new levels of transistor parallelism. Figure 1 looks at the past and present, and depicts the different levels of parallelism (*y*-axis) that have emerged as computing evolved over the decades (*x*-axis).

In Figure 1, Bit-level parallelism (BLP) refers to performing basic operations (arithmetic, and so forth) in parallel. It was common in early computers and was later enhanced with larger word sizes in commodity systems. Instruction-level parallelism (ILP) is the execution of logically sequential instructions concurrently with pipelining, superscalar, and increasing speculation. Thread-level parallelism (TLP) is the use of multiple processor cores, which initially started with discrete processors and were later integrated as on-chip cores. Data-level parallelism (DLP) pertains to performing similar operations on multiple data operands via arrays and pipelines that achieved broad success via general-purpose graphics processing units (GP-GPUs).

In this Viewpoint and in Figure 1, we assert that another major parallelism level is emerging: Accelerator-Level Parallelism (ALP). We define **ALP** *as the parallelism of workload components concurrently executing on multiple accelerators*. A goal of ALP is to unlock many accelerators at the same time in a manner analogous to how ILP concurrently employs multiple functional units. ALP does not replace other parallelism levels but builds upon them, as most accelerators internally employ one or more of BLP, ILP, TLP, and DLP. Moreover, much like ILP that has been exploited at different levels of the stack, ranging from superscalar and out-of-order execution at the microarchitecture level up to instruction scheduling at the compiler level, ALP opens up many degrees of freedom for novel hardware and software design and optimization. It also opens up possibilities for new runtime resource management, which is analogous to heterogeneous scheduling across CPUs and GPUs, but with the added complexity of scheduling tasks in real time across a sea of hardware accelerators.

ALP is emerging today. Modern chipsets for mobile, edge, and cloud computing are beginning to concurrently employ multiple accelerators. We next present a case study of ALP in mobile SoCs to understand how ALP is currently used, albeit in a somewhat limited form, and then lay a foundation for future work that can exploit ALP more generally.

## Mobile SoCs as Harbingers of Multiple Accelerators Using ALP

Driven by the need for extreme energy efficiency, mobile SoCs are the very early adopters of ALP. For SoCs from four major vendors—Apple, Qualcomm, Samsung, and Huawei—much less than 50% of the die is dedicated to the CPUs, as shown in the image on the first page of this Viewpoint. The majority of the area is dedicated to specialized accelerators, such as a Digital Signal Processor, Image Signal Processor, GPU, Neural Processing Unit, and Video Encoder/Decoder, as well as I/O interfaces for audio, networking, video.

It is common in smartphone SoCs for workloads to exhibit ALP with multiple accelerators in concurrent—not exclusive—use. Figure 2 shows a 4K, 60 frame-per-second video capture use case with two paths. One path goes to the display, rendering real-time content to the end user, and the other path goes to flash storage to save the content for offline viewing. In this example, data traverses accelerators with both parallelism (two paths) and pipelining, all choreographed by CPUs (not shown). In other use cases like an interactive multiparty videoconferencing application, data flow, and CPU choreographing can be even more dynamic and complex. Nevertheless, we expect accelerators to increasingly handle "data plane" computation while CPUs retain the "control plane" tasks. Doing so will enable richer computation from a fixed power budget, valuable from smartphones to cars to the cloud.

**Figure 1. A snapshot of parallelism over the years, showing how the various forms of parallelism were exploited through different types of architectural mechanisms.**

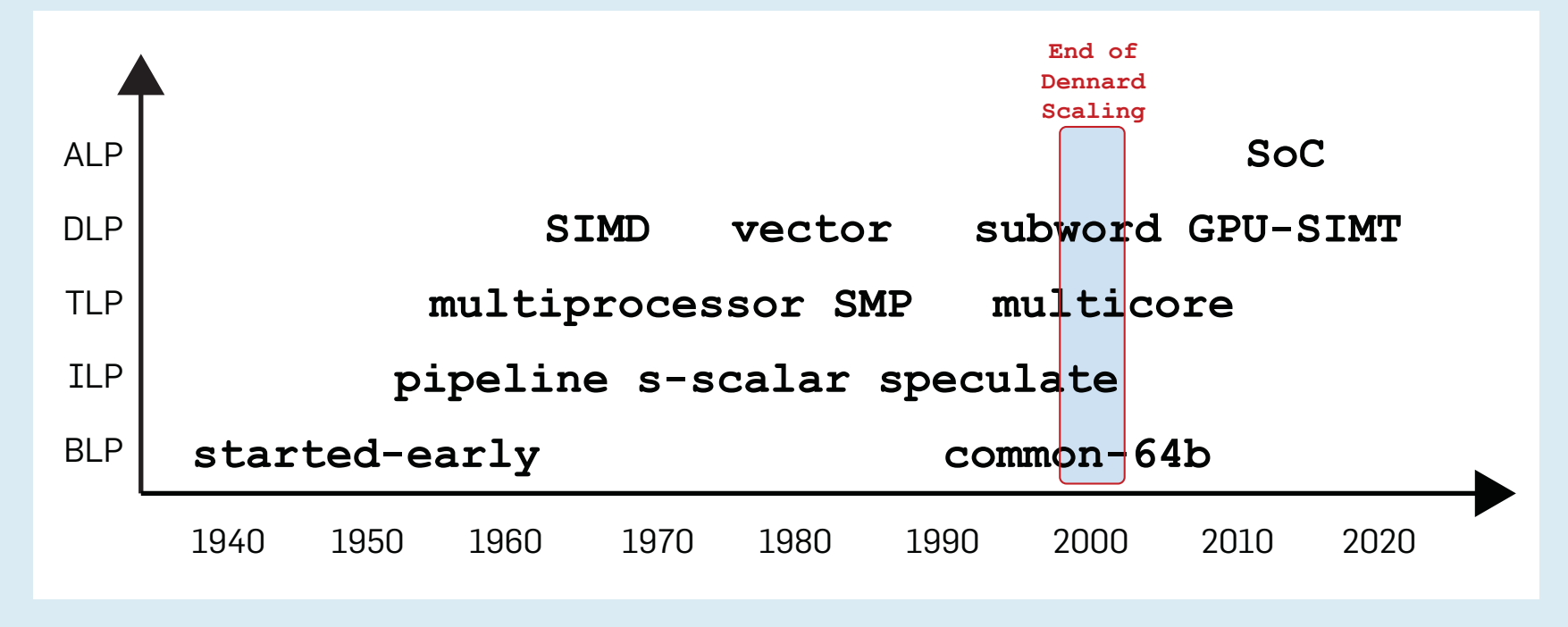

**Figure 2. ALP in action in a 4K video capture use case on a smartphone.[7]**

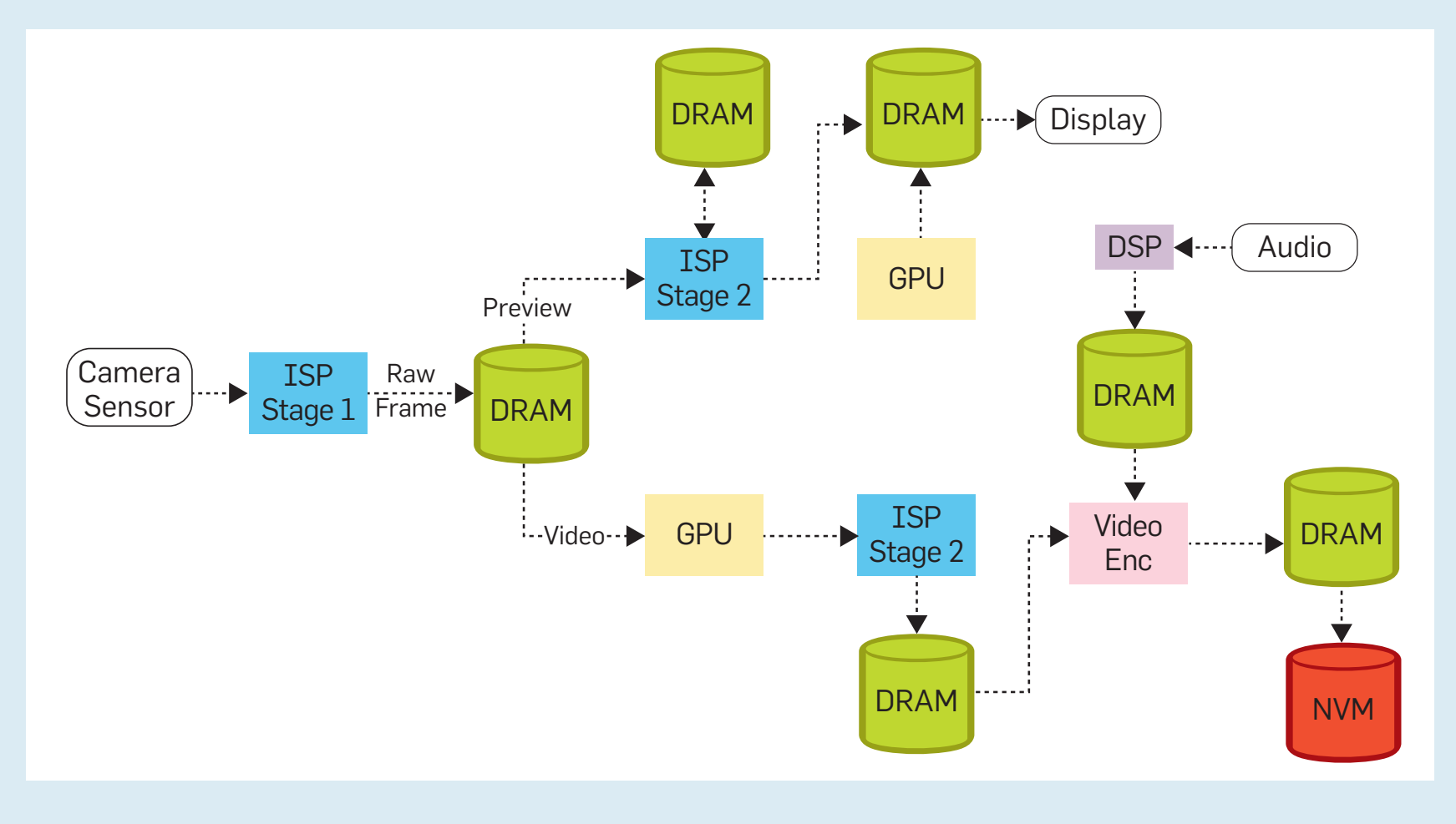

Mobile SoCs are clearly relying on ALP for low-power and efficient execution. However, they are not yet exploiting the full potential of ALP, which we see as needed for recouping the flexibility that the CPU delivered for decades. For instance, in the above example, the dataflow and the binding between the application tasks and accelerators is fixed. The ISP cannot be programmatically repurposed for tasks aside from processing camera image inputs. To this end, we believe we need better science and engineering toward ALP utilization.

## Toward a Science for Multiple-Accelerator Systems Using ALP

John Hennessy and David Patterson asserted in their 2018 Turing Award Lecture that we are upon a new golden age for computer architecture.[3] We assert that the challenge put forth by Hennessy and Patterson ought to be generalized to a new golden age for computer science and engineering and that employing multiple accelerators with ALP is an opportunity that opens up new vistas for research as accelerators are integrated into complex SoCs. We do not know all of the possibilities, but we discuss some ideas here to seed research directions.

A key challenge is developing abstractions and implementations to enable programmers to target the whole SoC and implementers to holistically design its software and hardware. We take inspiration from the Single Instruction Multiple Thread (SIMT) model that effectively abstracts GPU hardware's cornucopia of parallelism and scheduling mechanisms. SIMT both enabled GPUs to expand from graphics workloads to general-purpose DLP use and enabled software-hardware implementation improvements beneath the abstraction.

As ALP emerges, we expect new paradigms must be invented to flexibly and effectively exploit its potential. This is not the case today. In contrast to a SIMT-like holistic view, today's SoCs only exploit ALP in limited niches with each accelerator acting as a "silo" with its own programming model, and often its own (domain-specific) language, runtime, software development kit (SDK), and driver interface. While employing multiple accelerators with no abstraction can work in restricted situations (for example, for 10–20 phone use cases), it is unlikely to make ALP generally useful. How can we transcend per-accelerator software silos of different languages, SDKs, and so forth? What are abstractions and mechanisms for scheduling/sequencing accelerators or partitioning/virtualizing them (perhaps stream data flow)? What belongs in runtimes versus above/below the OS hardware abstraction layer?

Even more than previously parallel levels, ALP exploitation will likely require software-hardware co-design due to the heterogeneous nature of accelerators and ALP. Moreover, this is also likely to incentivize computer-aided design tool chain innovations to facilitate the rapid exploration of heterogeneous design spaces. ALP implementations should aspire toward globally optimal software-hardware systems, whereas much good work today focuses on making each accelerator "locally" optimal. While good accelerators are essential, a collection of locally optimal accelerators is unlikely to be globally optimal. For this reason, we need better models[4] and methods for holistically designing SoCs from accelerator, memory, and interconnect components, more like how processor cores are crafted from ALUs, register files, and buses. Analysis in both cases centers on parallel operation: ALP for SoCs and ILP for cores.

In more detail, there are many ALP questions that need better answers and better methods for systematically determining answers. For instance, from a compute perspective, we lack the fundamental science on how we must select, size, make efficient, and sometimes combine similar accelerators? Similarly, from a memory perspective, when should on-chip memory be private to accelerators or shared? When should this memory be a software-visible scratchpad or software-transparent cache? From an integration perspective, how do we best communicate data (shared memory or queues) and control (polling, interrupts, other) among accelerators? From an operational perspective, once an SoC is deployed, can we schedule heterogeneous parallel resources with (non-convex) optimization or must heuristics suffice? In sum, a more systematic approach is needed to design many accelerators as blocks to create holistic ALP systems that excel at performance and cost goals.

## Conclusion

This Viewpoint has argued that employing multiple accelerators with ALP has much promise for enhancing future computing efficiency, that we do not yet know how to do it well beyond niches, and that we can work together to make this happen. We have identified *what* the opportunity is, but leave to our readers *how* best to solve it. C

**Mark D. Hill** (markhill@cs.wisc.edu) is Hardware Partner Architect at Microsoft and Professor Emeritus at the University of Wisconsin-Madison, Madison, WI, USA.

**Vijay Janapa Reddi** (vj@eecs.harvard.edu) is an associate professor in the John A. Paulson School of Engineering and Applied Sciences (SEAS) at Harvard University, Cambridge, MA, USA.

Mark D. Hill contributed to this work before joining Microsoft.